\documentclass[12pt]{article}
\usepackage{graphicx}
\usepackage{cite}

\def \bea{\begin{eqnarray}}
\def \beq{\begin{equation}}
\def \b{{\cal B}}

\def \eea{\end{eqnarray}}
\def \eeq{\end{equation}}

\def \3half{\frac{3}{2}}

\def \ket#1{|#1 \rangle}

\def \obs{\overline{B}_s}

\def \vr{\vec{\rho}_T}
\def \vrp{\vec{\rho}_H}

\begin{document}
\begin{flushright}
EFI 08-23 \\
August 2008 \\
arXiv:0808.3761 \\
\end{flushright}
\centerline{\bf Flavor symmetry for strong phases and determination of 
$\beta_s, \Delta\Gamma$ in $B_s \to J/\psi \phi$}
\bigskip
\centerline{Michael Gronau\footnote{On sabbatical leave from the Physics 
Department, Technion, Haifa 32000, Israel.} and Jonathan L. Rosner}
\medskip
\centerline{\it Enrico Fermi Institute and Department of Physics,
 University of Chicago} 
\centerline{\it Chicago, IL 60637, U.S.A.} 
\bigskip
\begin{quote}
Studies of angular and time distributions and CP-violating observables in $B_s
\to J/\psi \phi$ decays yield a space of parameters which can be considerably
reduced if relative strong phases among different amplitudes are specified.
We show that the relations between $B_s \to J/\psi \phi$ and $B^0 \to
J/\psi K^{*0}$ amplitudes given by flavor symmetry [actually U(3) rather
than SU(3)] are likely to be quite reliable, and hence the use of
strong phases from $B^0 \to J/\psi K^{*0}$ in the analysis of $B_s \to J/\psi
\phi$ is justified.  We point out the potential advantage
of using helicity angles over transversity angles, and comment on a way to 
measure a sizable CP-violating phase  independent of strong phases.
\end{quote} 

\leftline{\qquad PACS codes:  12.15.Hh, 12.15.Ji, 13.25.Hw, 14.40.Nd} 

\medskip
The decay $B_s \to J/\psi \phi$ is of great interest in the study of
CP violation.  In the Standard Model in which CP violation is generated
purely by phases in the Cabibbo-Kobayashi-Maskawa (CKM) matrix, the CP
asymmetry in this process is expected to be small, as it should be governed
by the $B_s$--$\obs$ mixing phase $\phi_M = - 2 \beta_s$, where $\beta_s = {\rm
Arg}(- V_{ts}V^*_{tb}/V_{cs}V^*_{cb}) = \lambda^2 \eta \simeq 0.02$
\cite{Wolfenstein:1983yz}, with $\lambda = 0.2255 \pm 0.0019$ \cite{PDG} and
$\eta \simeq 1/3$ parameters in the CKM matrix.

Recently both the CDF Collaboration \cite{Aaltonen:2007he} and
the D0 Collaboration \cite{Abazov:2008fj}, working at the Fermilab Tevatron
collider, have reported fits to angular and time distributions
of flavor-tagged $B_s \to J/\psi \phi$ decays which favor larger values
of $2\beta_s$.  (Earlier reports by CDF and D0 which did not apply flavor
tagging can be found in Refs.\,\cite{Aaltonen:2007gf,Acosta:2004gt} and
\cite{Abazov:2007tx}.) In the absence of information about relative strong
phases between different amplitudes, these solutions have a two-fold ambiguity
which can be eliminated with strong phase information \cite{Dunietz:2000cr}
from the decay $B^0 \to J/\psi K^{*0}$ using flavor symmetry.  The D0 analysis
uses this information, while CDF finds consistency with it.

The present note examines the robustness of this assumption, and finds it to
be valid to a sufficient degree that the two-fold ambiguity indeed can be
eliminated.  Our approach
is the inverse of that of Ref.\ \cite{Dighe:1998vm}, which suggested making use
of strong phase information in $B_s \to J/\psi \phi$ (assuming Standard
Model weak phases) to resolve a discrete ambiguity in the sign of $\cos 2
\beta$ in $B^0 \to J/\psi K^{*0}$.
In passing we make some remarks about the angular
and time dependence of the $B_s \to J/\psi \phi$ observables.  
 In order to avoid a potential bias due to
an uncertainty in angular acceptance, 
we suggest using a set of angles differing from those used by CDF and D0.
We offer the possibility that a large value of $2\beta_s$ could show up
in time oscillations which do not depend on strong phases.

In order to specify the two-fold ambiguity resolved by strong phases,
we first review the angular and time dependence in decays $M \to V_1 V_2$,
where $M$ is a spinless meson, $V_{1,2}$ are vector mesons, and $V_1 \to \ell^+
\ell^-$, $V_2 \to P_A P_B$.  Here $\ell^+$ and $\ell^-$ are leptons, while
$P_A$ and $P_B$ are pseudoscalar mesons.  We use the {\it transversity basis}
\cite{Dunietz:1990cj} and its application to these decays in Ref.\
\cite{Dighe:1995pd}.

In the rest frame of $V_1$ let the direction of $V_2$ define the $x$ axis. Let
the plane of the $P_A P_B$ system define the $y$ axis, with $p_y(P_A) > 0$, so
the normal to that plane (taking a right-handed coordinate system) defines the
$z$ axis.  For $\phi \to K^+ K^-$ and $K^{*0} \to K^+ \pi^-$, we shall take
$P_A \equiv K^+$.  A unit vector $n$ in the direction of the $\ell^+$ in $V_1$
decay may be defined to have components
\beq
(n_x, n_y, n_z) =
 (\sin \theta_T \cos \phi_T, \sin \theta_T \sin \phi_T, \cos \theta_T)~,
\eeq
thereby defining the polar and azimuthal transversity angles $\theta_T$ and
$\phi_T$.  
A third angle $\psi$ is defined as that of $P_A$ in the $V_2$ rest frame
relative to the helicity axis (the negative of the
direction of $V_1$ in that frame).

The decay of a spinless meson $M$ to $V_1 V_2$ is characterized by three
independent amplitudes \cite{Rosner:1990xx}, corresponding to linear
polarization states of the vector mesons which are either longitudinal (0), or
transverse to their directions of motion and parallel ($\parallel$) or
perpendicular ($\perp$) to one another.  The states $0$ and $\parallel$ are
P-even, while the state $\perp$ is P-odd.  When $V_1$ and $V_2$ are eigenstates
of C with the same eigenvalue (as in the case $V_1 = J/\psi,~V_2 = \phi$), the
properties under P are the same as those under CP.  We denote the invariant
amplitudes by $A_0,~A_\parallel,~A_\perp$, using the normalization
\beq
\label{normal}
|A_0|^2 + |A_\parallel|^2 + |A_\perp|^2 = 1~~.
\eeq
The corresponding distribution in $\vr \equiv (\theta_T,\phi_T,\psi)$ is
\cite{Dighe:1995pd}
$$
\frac{d^4 \Gamma [M \to (\ell^+\ell^-)_{V_1} (P_A P_B)_{V_2}]}
{d \cos \theta_T~d \phi_T~d \cos \psi~dt} \propto \frac{9}{32 \pi}
[|A_0|^2 f_1(\vr) + |A_\parallel|^2 f_2(\vr) + |A_\perp|^2 f_3(\vr)
$$
\beq
+ {\rm Im}(A_\parallel^* A_\perp) f_4(\vr) + {\rm Re}(A_0^* A_\parallel)
f_5(\vr) + {\rm Im}(A_0^* A_\perp) f_6(\vr)]~,
\label{eqn:angdep}
\eeq
where
$$
f_1(\vr) \equiv 2 \cos^2 \psi (1 - \sin^2 \theta_T \cos^2 \phi_T)~,~~
f_2(\vr) \equiv \sin^2 \psi (1 - \sin^2 \theta_T \sin^2 \phi_T)~,
$$
$$
f_3(\vr) \equiv \sin^2 \psi \sin^2 \theta_T~,~~
f_4(\vr) \equiv - \sin^2 \psi \sin 2 \theta_T \sin \phi_T~,
$$
\beq \label{eqn:fdef}
f_5(\vr) \equiv \frac{1}{\sqrt{2}}\sin 2 \psi \sin^2 \theta_T \sin 2 \phi_T~,~~
f_6(\vr) \equiv \frac{1}{\sqrt{2}}\sin 2 \psi \sin 2 \theta_T \cos \phi_T~.
\eeq

We now describe the time-dependence of the decay for $M = B_s$, $V_1 = J/\psi$,
$V_2 = \phi$, following discussions in Refs.\ \cite{Fleischer:1996aj,%
Dighe:1998vk,Dunietz:2000cr,CDF9458}. 
The mass-eigenstate combinations of $B_s$ and $\obs$, which have definite
time-development properties, are
\beq
\ket{B_{sL}} = p \ket{B_s} + q \ket{\obs}~,~~
\ket{B_{sH}} = p \ket{B_s} - q \ket{\obs}~,
\eeq
with $|p|^2 + |q|^2 = 1$.
In the limit in which $|\Delta \Gamma/ \Delta m| \ll 1$ (a good approximation
here), $q/p \simeq e^{2 i \beta_s}$, a pure phase.  Heavy and light
eigenstate masses and widths are $(m_H,m_L)$ and $(\Gamma_H,\Gamma_L)$; their
differences are $\Delta m_s \equiv m_H - m_L$, $\Delta \Gamma \equiv \Gamma_L -
\Gamma_H$ (note the sign).  The average of CDF and D0 measurements gives
$\Delta m_s = 17.78 \pm 0.12$ ps$^{-1}$ \cite{Barberio:2008fa}.
In the Standard Model limit of small CP violation in CKM-favored $B_s$ decays,
the decay width of the CP-even lighter state $L$ is greater than that of the
CP-odd heavier state $H$.  This property, $\Delta\Gamma >0$, has been shown
some time ago by explicit calculations of decays to CP-even and CP-odd
eigenstates \cite{Aleksan:1993qp}.  A recent calculation obtains $\Delta
\Gamma = 0.096 \pm 0.039$ ps$^{-1}$ \cite{Lenz:2006hd}.  The
average $\Gamma \equiv (\Gamma_L + \Gamma_H)/2$ is $1/\bar \tau(B_s) =
1/(1.478^{+0.020}_{-0.022} {\rm~ps})$ \cite{Barberio:2008fa}.

Defining a parameter $\eta \equiv +1$ for a tagged $B_s$ and $-1$ for a tagged
$\obs$, we may write 
$$
\frac{d^4 \Gamma [B_s(\obs) \to (\ell^+\ell^-)_{J/\psi} (K^+ K^-)_{\phi}]}
{d \cos \theta_T~d \phi_T~d \cos \psi~dt} \propto \frac{9}{32 \pi} \{
[|A_0|^2 f_1(\vr) +|A_\parallel|^2 f_2(\vr)]{\cal T}_+
 + |A_{\perp}|^2 f_3(\vr)~{\cal T}_-
$$
\beq
\label{angles+time}
+ |A_\parallel||A_{\perp}| f_4(\vr)~{\cal U} + |A_0||A_\parallel|
\cos(\delta_\parallel) f_5(\vr)~{\cal T}_+ + |A_0||A_{\perp}| f_6(\vr)~
{\cal V} \}~~.
\eeq
Here $A_i\equiv A_i(t=0)$, while dependence on time is given by the four
functions
\beq \label{eqn:T}
{\cal T}_\pm \equiv e^{-\Gamma t}[\cosh(\Delta \Gamma t/2) \mp \cos(2 \beta_s)
\sinh(\Delta \Gamma t/2) \mp \eta \sin(2 \beta_s) \sin(\Delta m_s t)]~,
\eeq
$$
{\cal U} \equiv e^{-\Gamma t}[\eta \sin(\delta_\perp - \delta_\parallel)
\cos(\Delta m_s t) - \eta \cos(\delta_\perp - \delta_\parallel) \cos(2 \beta_s)
\sin(\Delta m_s t)
$$
\beq \label{eqn:U}
+ \cos(\delta_\perp - \delta_\parallel) \sin(2 \beta_s) \sinh(\Delta \Gamma
 t/2)]~,
\eeq
$$
{\cal V} \equiv e^{-\Gamma t}[\eta \sin(\delta_\perp) \cos(\Delta m_s t)
 - \eta \cos(\delta_\perp) \cos(2 \beta_s) \sin(\Delta m_s t)~~~~~~~~~~~~~~
$$
\beq \label{eqn:V}
+ \cos(\delta_\perp) \sin(2 \beta_s) \sinh(\Delta \Gamma t/2)]~,~~~~~
\eeq
where relative strong phases are defined by
\beq
\delta_\parallel \equiv {\rm Arg}(A_\parallel(0) A^*_0(0))~,~~
\delta_\perp \equiv {\rm Arg}(A_\perp(0) A^*_0(0))~.
\eeq

\begin{table}
\caption{Comparison of BaBar (Ba) \cite{Aubert:2007hz}, Belle (Be)
\cite{Itoh:2005ks},
Heavy Flavor Averaging Group (HFAG) averages of these
values (H) \cite{HFAGAv}, and CDF (C) \cite{CDF8950} magnitudes and phases of
$B^0 \to J/\psi K^{*0}$ decay amplitudes.  The CDF values are not included
in the HFAG averages.
\label{tab:comp}}
\begin{center}
\begin{tabular}{c c c c c} \hline \hline
 & $|A_0|^2$ & $|A_\parallel|^2$ & $\delta_\parallel$ & $\delta_\perp$\\ \hline
Ba & 0.556$\pm$0.009$\pm$0.010 & 0.211$\pm$0.010$\pm$0.006
 & --2.93$\pm$0.08$\pm$0.04 & 2.91$\pm$0.05$\pm$0.03 \\
Be & 0.574$\pm$0.012$\pm$0.009 & 0.231$\pm$0.012$\pm$0.008
 & --2.89$\pm$0.09$\pm$0.01 & 2.94$\pm$0.06$\pm$0.01 \\
H & 0.56$\pm$0.01 & 0.219$\pm$0.009 & --2.91$\pm$0.06 & 2.92$\pm$0.04 \\
C & 0.569$\pm$0.009$\pm$0.009 & 0.211$\pm$0.012$\pm$0.006
 & --2.96$\pm$0.08$\pm$0.03 & 2.97$\pm$0.06$\pm$0.01 \\ \hline \hline
\end{tabular}
\end{center}
\end{table}

We now can see the source of the discrete two-fold ambiguity. 
The $\cos\delta_{\parallel}$ term in the decay distribution and the 
time-dependent functions ${\cal T}_\pm,~{\cal U},~{\cal V}$
are invariant under the simultaneous substititions~\cite{Aaltonen:2007he}
\beq
2 \beta_s \to \pi - 2 \beta_s~,~~\Delta \Gamma \to - \Delta \Gamma~,
~~\delta_\parallel \to - \delta_\parallel~,~~
\delta_\perp \to \pi - \delta_\perp~.
\eeq
Without tagging information [i.e., setting $\eta = 0$ in Eqs.\
(\ref{eqn:T}--\ref{eqn:V})], there would be additional invariances under
$\beta_s \to - \beta_s,~\delta_\parallel \to - \delta_\parallel,~
\delta_\perp \to \pi - \delta_\perp$ and under $2 \beta_s \to 2 \beta_s-\pi~,
\Delta \Gamma \to - \Delta \Gamma$.

Magnitudes and phase amplitudes for $B^0 \to J/\psi K^{*0}$ obtained by BaBar
\cite{Aubert:2007hz}, Belle \cite{Itoh:2005ks}, the Heavy Flavor Averaging
Group (HFAG) averages of these values \cite{HFAGAv}, and CDF \cite{CDF8950} are
compared in Table \ref{tab:comp}.
These entail $\sin(\delta_\perp - \delta_\parallel) \simeq -0.4$,
$\cos(\delta_\perp - \delta_\parallel) \simeq 0.9$,
$\sin(\delta_\perp) \simeq 0.2$, $\cos(\delta_\perp) \simeq -0.98$.
A change in sign of $\cos(\delta_\perp - \delta_\parallel)$ and
$\cos(\delta_\perp)$ has nearly maximal effect on
the $\sin(\Delta mt)$ and $\sinh(\Delta\Gamma t/2)$ terms in 
${\cal U}$ and ${\cal V}$ if not balanced by changes in sign
of $\cos(2 \beta_s)$ and $\Delta \Gamma$.  Hence such strong phases, if
employed in fits to $B_s \to J/\psi \phi$ tagged time-dependent decays,
will be effective in resolving the twofold ambiguity
in $\beta_s$ and $\Delta\Gamma$. It is notable that
magnitudes of amplitudes close to those quoted in Table
\ref{tab:comp}, and solutions for $\delta_{\parallel}$ and $\delta_\perp$ in  
ranges of values consistent with those in Table \ref{tab:comp}, were obtained 
by the CDF Collaboration in a fit to {\it untagged} $B_s \to J/\psi \phi$ 
decays ~\cite{Aaltonen:2007gf,CDF8950} and by the D0 Collaboration in a {\it flavor-tagged}
study of this process~\cite{Abazov:2008fj}.

\begin{figure}
\mbox{\includegraphics[width=0.48\textwidth]{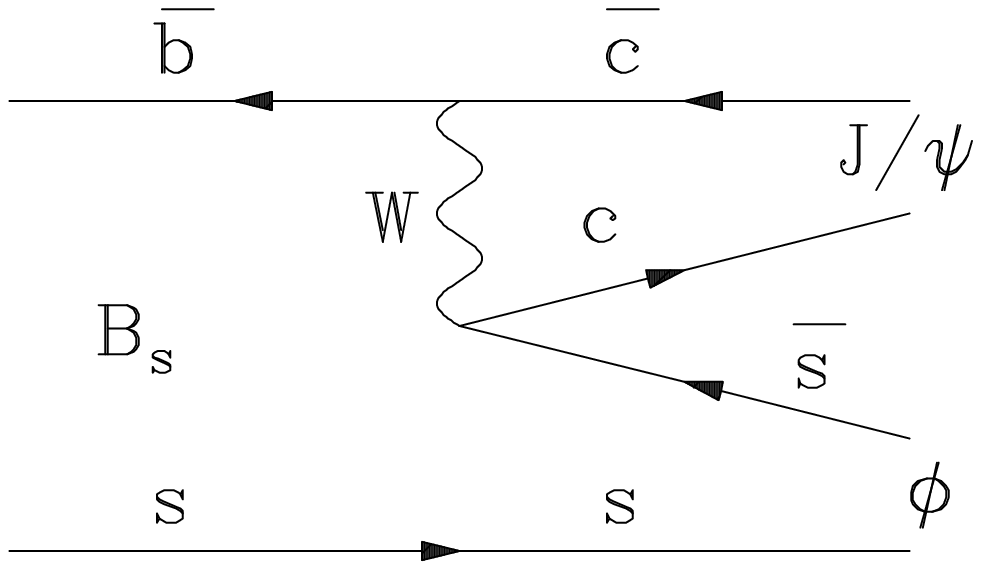}
      \includegraphics[width=0.48\textwidth]{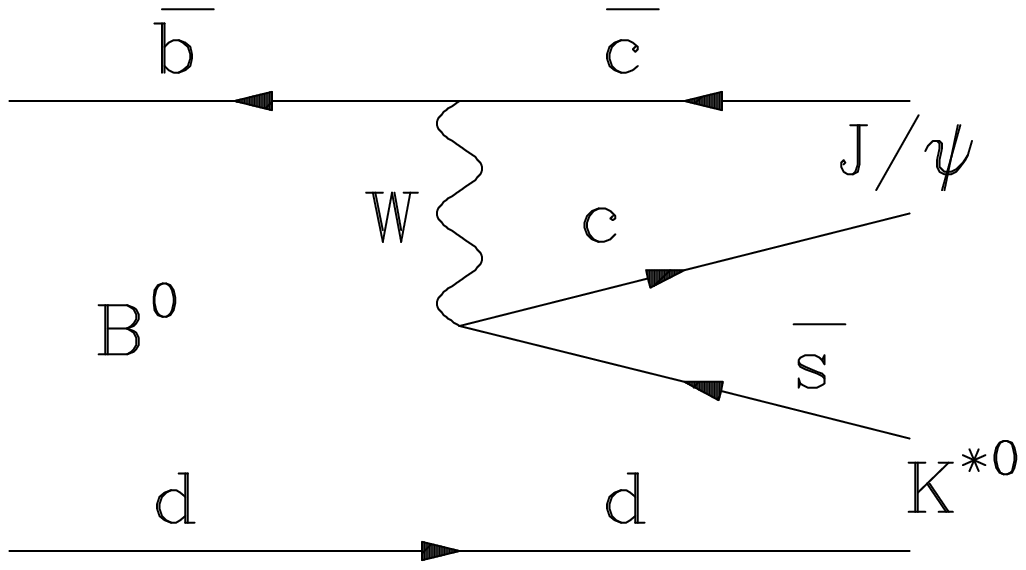}}
\caption{Color-suppressed tree diagrams, providing the dominant contribution
to $B_s \to J/\psi \phi$ (left) and $B^0 \to J/\psi K^{*0}$ (right)
\label{fig:cs}}
\end{figure}

We now discuss our reason for expecting strong phases in $B_s \to J/\psi \phi$
very similar to those in $B^0 \to J/\psi K^{*0}$.  Our conclusion is based
on the high degree of similarity governing these two processes.  They are
dominated by the color-suppressed tree diagrams illustrated in Fig.\
\ref{fig:cs}.  These two processes differ only by the substitution
$s \leftrightarrow d$ of the spectator quark.  Indeed, they are characterized
by similar branching ratios \cite{PDG}:
\beq
\b(B_s \to J/\psi \phi) = (0.93 \pm 0.33) \times 10^{-3}~,~~
\b(B^0 \to J/\psi K^{*0}) = (1.33 \pm 0.06) \times 10^{-3}~.
\eeq
(The former is based on a very old value \cite{Abe:1996kc} and deserves to be
updated.) Taking account of the ratio of lifetimes \cite{Barberio:2008fa}
$\tau(B_s)/\tau(B^0)=0.966\pm 0.015$, one then finds the ratio of decay rates
to be
\beq\label{ratio-of-widths}
\frac{\Gamma(B^0 \to J/\psi K^{*0})}{\Gamma(B_s \to J/\psi \phi)}
= 1.38 \pm 0.49~~.
\eeq

\begin{figure}
\mbox{\includegraphics[width=0.48\textwidth]{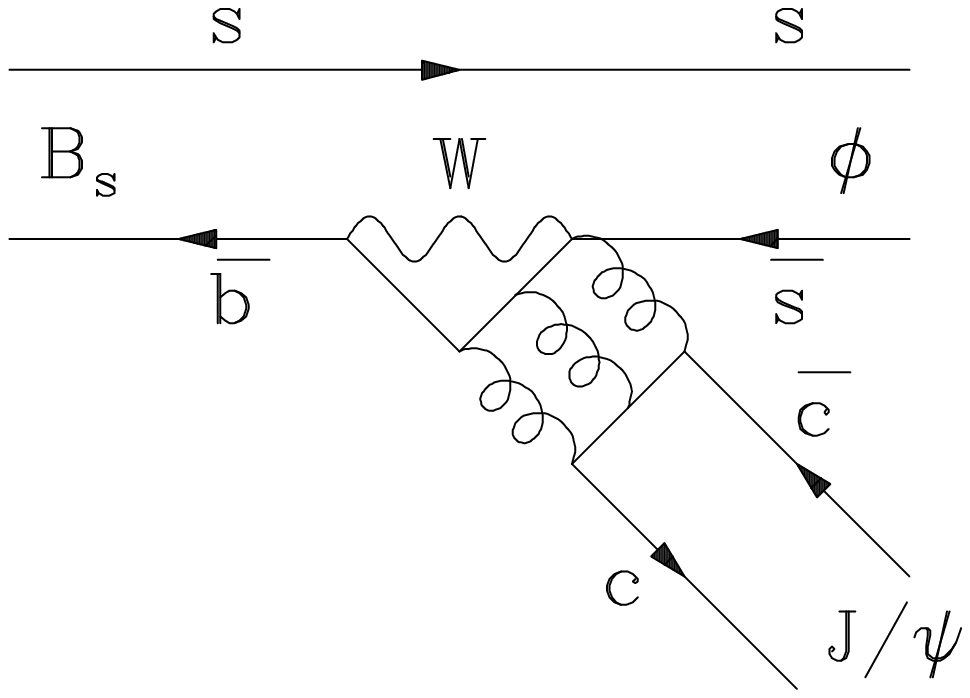}
      \includegraphics[width=0.48\textwidth]{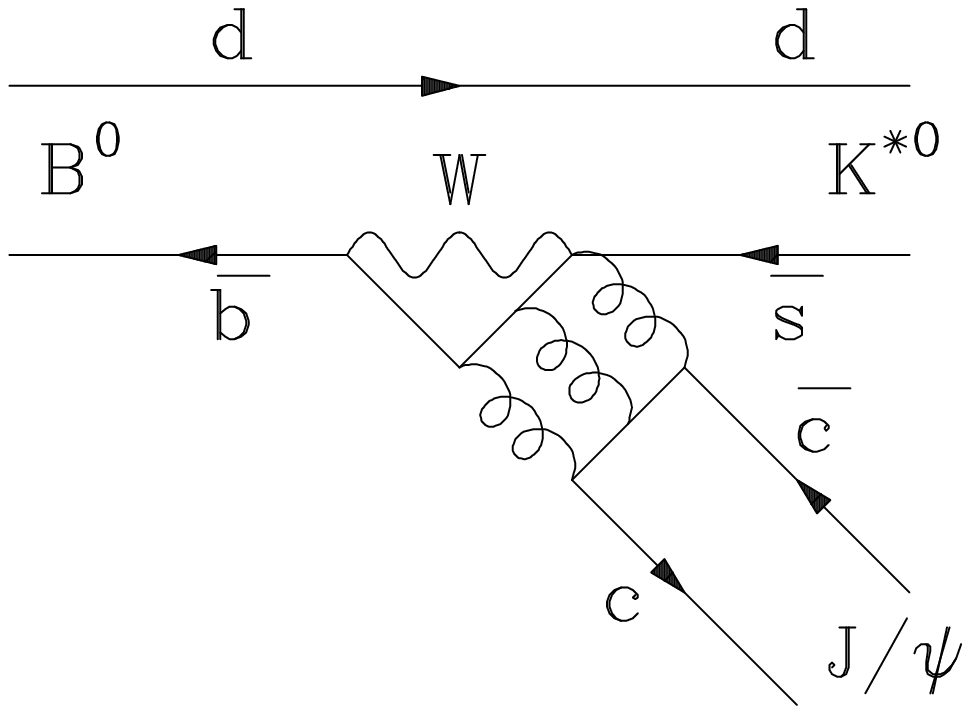}}
\caption{Gluonic penguin diagrams contributing
to $B_s \to J/\psi \phi$ (left) and $B^0 \to J/\psi K^{*0}$ (right)
\label{fig:p3g}}
\end{figure}

\begin{figure}
\mbox{\includegraphics[width=0.48\textwidth]{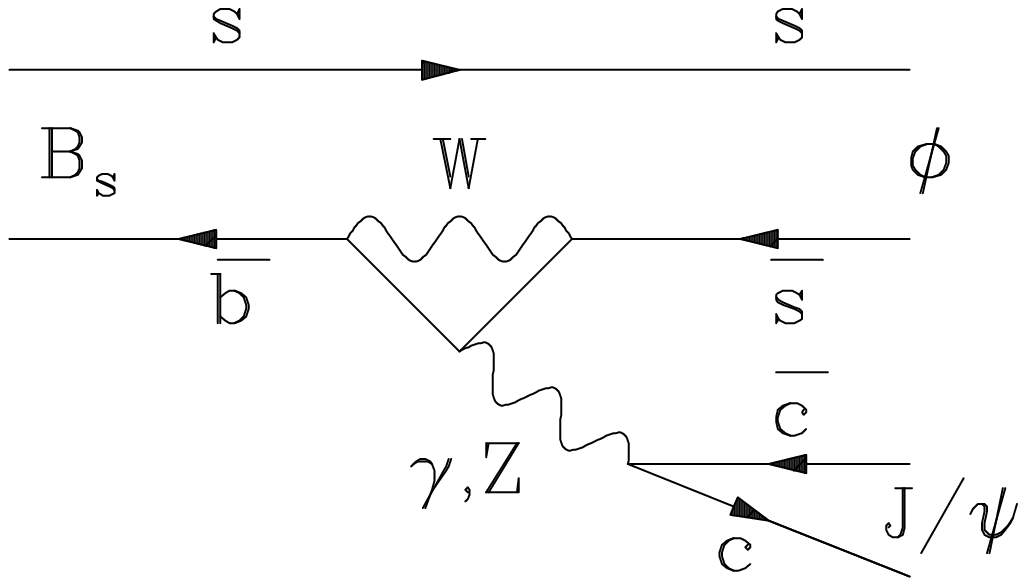}
      \includegraphics[width=0.48\textwidth]{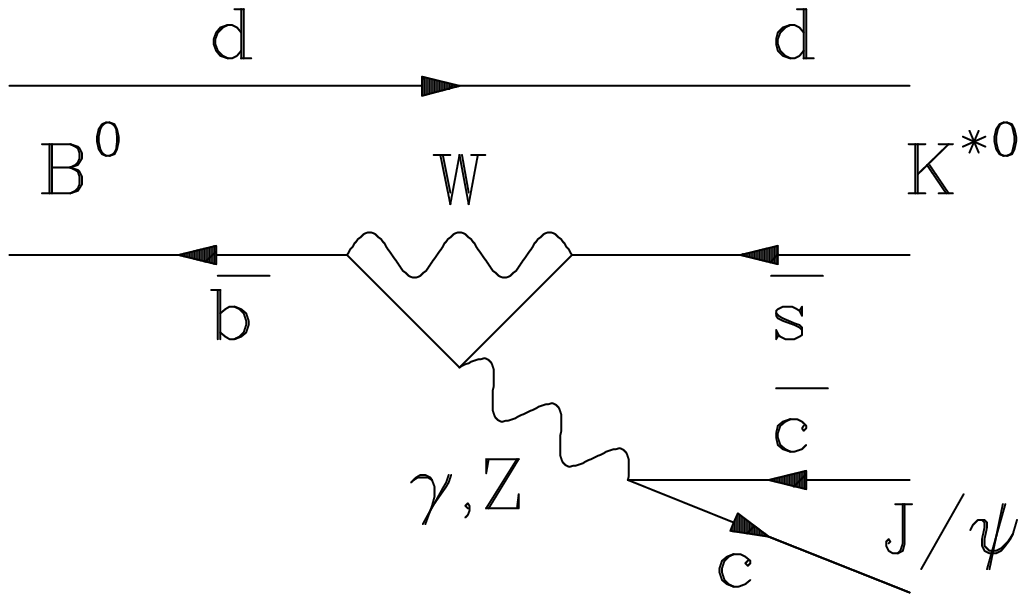}}
\caption{Electroweak penguin diagrams contributing
to $B_s \to J/\psi \phi$ (left) and $B^0 \to J/\psi K^{*0}$ (right).
\label{fig:ewp}}
\end{figure}

Additional processes contributing to both decays and differing only by the
substitution $s \leftrightarrow d$ of the spectator quark involve gluonic
and electroweak penguin amplitudes, illustrated respectively in Figs.\
\ref{fig:p3g} and \ref{fig:ewp}.
One expects that the degree of flavor symmetry violation associated with
the coherent sum of Figs.\ \ref{fig:cs}--\ref{fig:ewp} will not be greater
than the flavor violation in the individual components.  Typical violations
of this symmetry do not exceed 30\% in the magnitudes of amplitudes.  

Experimental evidence for approximate SU(3) invariance of both the 
magnitudes of amplitudes and their relative strong phases is provided by 
decay rates and CP asymmetries measured for $B$ meson decays into 
$\pi\pi, K\pi$ and $K\bar K$~\cite{Chiang:2004nm}. 
One particular test, which is sensitive to SU(3) invariance of relative 
strong phases, relates CP asymmetries in $B^0\to K^+\pi^-$ 
and $B^0\to \pi^+\pi^-$~\cite{Deshpande:1994ii,Gronau:1995qd},
\beq
\frac{A_{CP}(B^0\to K^+\pi^-)}{A_{CP}(B^0\to \pi^+\pi^-)} = 
-\frac{\b(B^0\to\pi^+\pi^-)}{\b(B^0\to K^+\pi^-)}~.
\eeq
Experimentally the two ratios read~\cite{Barberio:2008fa}
\beq
-0.255\pm 0.057 = -0.266\pm 0.014~.
\eeq
This shows that strong phases and not only magnitudes of amplitudes 
are approximately equal for SU(3)-related processes.


\begin{figure}
\mbox{\includegraphics[width=0.46\textwidth]{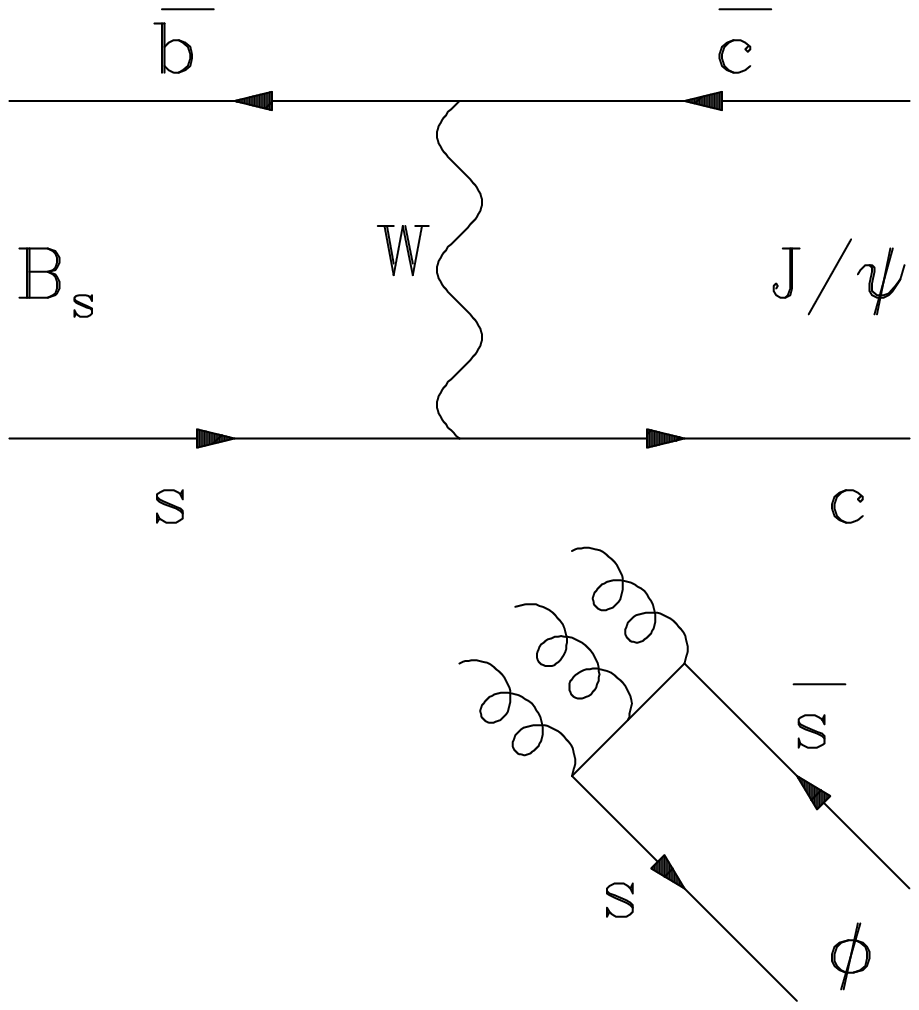} \hskip 0.2in
      \includegraphics[width=0.46\textwidth]{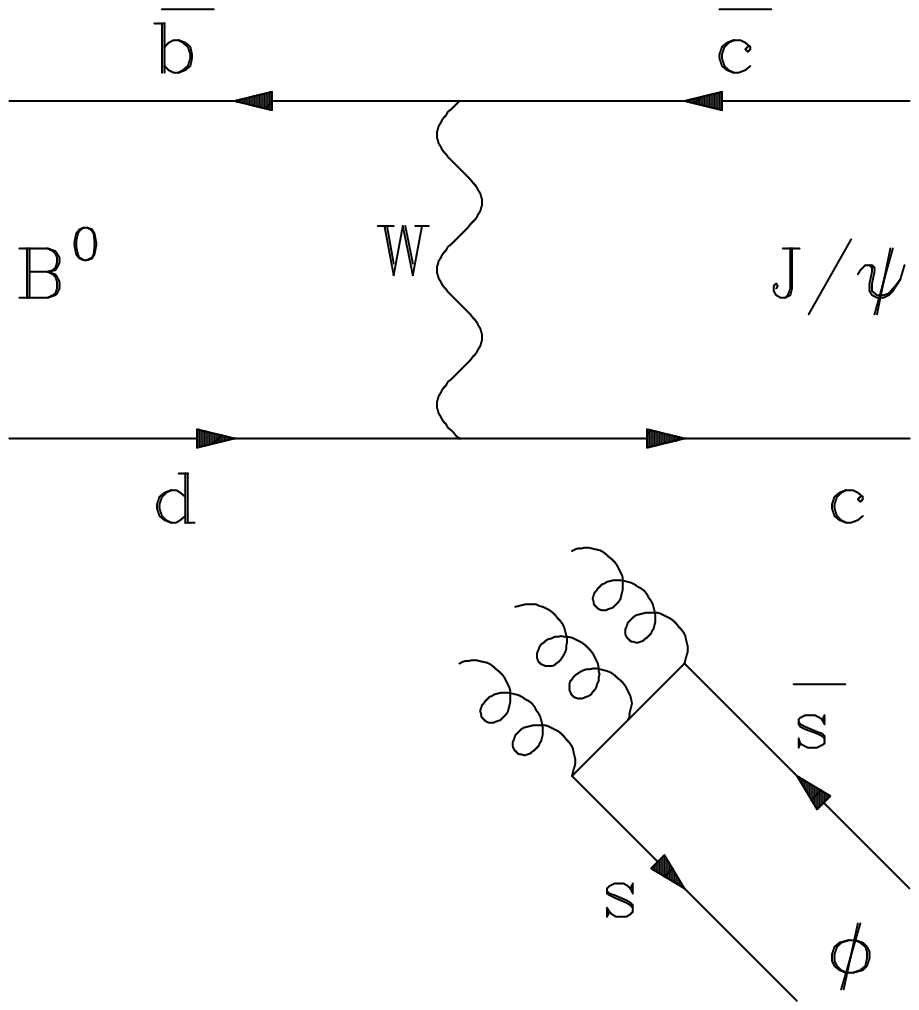}}
\caption{Left: exchange diagram contributing to $B_s \to J/\psi \phi$.  No
corresponding diagram can be drawn for $B^0 \to J/\psi K^{*0}$.  Right:
corresponding Cabibbo-suppressed diagram contributing to $B^0 \to J/\psi \phi$.
\label{fig:ex}}
\end{figure}

Special care must be taken when comparing $B_s\to J/\psi \phi$ and 
$B^0\to J/\psi K^{*0}$ using flavor SU(3). In fact, here one is not using SU(3)
but U(3), i.e., nonet symmetry.  One must investigate contributions
unique to the SU(3)-singlet component of the $\phi$.  The dominant diagram that
contributes to $B_s \to J/\psi \phi$ through the $\phi$ singlet component,
and not to $B^0 \to J/\psi K^{*0}$, is shown in Fig.\ \ref{fig:ex} (left).
Here the $\phi$ couples to the rest of the $W$-exchange diagram via a minimum
of three gluons, so this contribution is expected
to be suppressed by the Okubo-Zweig-Iizuka (OZI) \cite{OZI} rule.
Another diagram, doubly-OZI-suppressed, involves annihilation of the $b
\bar s$ pair to a $u,c,t$ loop (``penguin annihilation''), which is then
connected to the $c \bar c$ pair of the $J/\psi$ and the $s \bar s$ pair
of the $\phi$ each by a single gluon, with the $c \bar c$ and $s \bar s$
pair connected to each other by two more gluons.  We expect this diagram
to contribute even less than that in Fig.\ 4 (left)
because of an extra loop factor in the penguin annihilation diagram.

A corresponding Cabibbo-suppressed $W$-exchange diagram, shown in 
Fig.\ \ref{fig:ex} (right),
 (and a penguin annihilation diagram as mentioned above) can contribute to 
 $B^0 \to J/\psi \phi$, for which a new upper bound
is $\b[B^0 \to J/\psi \phi] < 9.4 \times 10^{-7}$ \cite{Liu:2008bt}.  We have
recently found this diagram to contribute a rate smaller than the prediction
$\b(B^0 \to J/\psi \phi) = (1.8 \pm 0.3) \times 10^{-7}$ due to
$\omega$--$\phi$ mixing \cite{Gronau:2008kk}.  We thus estimate that the
exchange process illustrated in Fig.\ \ref{fig:ex} (left) can contribute 
no more than $[(1 - \frac12 \lambda^2)^2/\lambda^2][\tau(B_s)/\tau(B^0)]
\b(B^0 \to J/\psi \phi)\simeq (3.2 \pm 0.5) \times 10^{-6}$ to $\b(B_s \to
J/\psi \phi)$.  This is only  about 1/300 of its measured value. 

We thus conclude that the similarity of 
amplitudes and strong phases for $B_s \to J/\psi \phi$
and $B^0 \to J/\psi K^{*0}$, at least to the degree needed to eliminate the
discrete ambiguity noted in Ref.\ \cite{Aaltonen:2007he}, is a well-founded
assumption.  The fact that the $\phi$ is a nonet, rather than octet, partner
of $K^{*0}$ is not an essential complication.

Because of the normalization (\ref{normal}), given magnitudes of transversity
amplitudes are equivalent to given values for their ratios. Symmetry 
breaking factors which are common to the three transversity amplitudes cancel 
in ratios of these amplitudes. Thus we expect symmetry relations between these
ratios in  $B_s \to J/\psi \phi$ and $B^0 \to J/\psi K^{*0}$ to hold 
more precisely than the equality $\sqrt{\Gamma(B_s \to J/\psi \phi)} \simeq
\sqrt{\Gamma(B^0 \to J/\psi K^{*0})}$, which 
may involve a U(3)-breaking correction up to the level of $30\%$.
Indeed, magnitudes of corresponding normalized transversity amplitudes measured for 
instance by the CDF Collaboration are equal within a few percent
 \cite{Aaltonen:2007gf,CDF8950},
\beq \label{magnitudes}
|A_0| =  \left\{ \begin{array}{c} 0.729\pm 0.015,~B_s\to J/\psi\phi~~\cr 
0.754 \pm 0.008,~B^0\to J/\psi K^*\end{array}\right.,~~~
|A_\parallel| =  \left\{ \begin{array}{c} 0.480\pm 0.029,~B_s\to
J/\psi\phi~~\cr 0.459 \pm 0.015,~B^0\to J/\psi K^* \end{array} \right.~,
\eeq
where the $B_s\to J/\psi\phi$ values are quoted assuming $\beta_s=0$.
We expect the relative strong phases $\delta_\parallel$ and $\delta_\perp$ in 
$B_s\to J/\psi\phi$ and $B^0\to J/\psi K^*$ to be equal  within a similar precision,
i.e., a few percent of $\pi$ or $\simeq 10^\circ$. Applying such an error when 
assuming equal strong phases for these two processes in fits to data 
would be a reasonable assumption.

We close with some remarks regarding angular and time distributions.

(1) All the analyses of $B_s \to J/\psi \phi$ and $B^0 \to J/\psi K^{*0}$ display 
appreciable dependence of acceptance on the transversity azimuthal angle $\phi_T$
\cite{Acosta:2004gt,Aaltonen:2007gf,Abazov:2007tx,CDF9458,Aubert:2007hz,CDF8950}.
This is because the angle between the leptons and the helicity axis, and hence
the fraction of events that will be selected by choosing leptons with
selection cuts on kinematic variables, depends on $\phi_T$. In contrast, if one
transforms the transversity angles $(\theta_T,\phi_T)$ to an equivalent set of
{\it helicity angles} $(\theta_H,\phi_H)$, one should expect the acceptance to
depend less on $\phi_H$, and all its variation due to lepton selection criteria
should be concentrated in $\theta_H$.  The angle $\phi_H$ (to be defined below)
describes the relative orientations of the decay planes defined by the decay
products of $V_1 = J/\psi$ and $V_2 = \phi$ and the helicity axis.
As rotations by $\phi_H$ only affect the relative orientations of these
planes, they should not lead to variations in efficiency of lepton selection.

To define the helicity angles we identify
\beq
(\sin \theta_T \cos \phi_T,~\sin \theta_T \sin\phi_T,~\cos \theta_T) =
(\cos \theta_H, \sin \theta_H \cos \phi_H, \sin \theta_H \sin \phi_H)~.
\eeq
The angular functions $g_j(\vrp) = f_j(\vr)$ for $\vrp\equiv(\theta_H,\phi_H,
\psi)$ then become
$$
g_1(\vrp) = 2 \cos^2 \psi \sin^2 \theta_H~,~~
g_2(\vrp) = \sin^2 \psi (1 - \sin^2 \theta_H \cos^2 \phi_H)~,
$$
$$
g_3(\vrp) = \sin^2 \psi (1 - \sin^2 \theta_H \sin^2 \phi_H)~,~~
g_4(\vrp) = - \sin^2 \psi \sin^2 \theta_H \sin 2 \phi_H~,
$$
\beq \label{eqn:fhel}
g_5(\vrp) = \frac{1}{\sqrt{2}}\sin 2 \psi \sin 2 \theta_H \cos \phi_H~,~~
g_6(\vrp) = \frac{1}{\sqrt{2}}\sin 2 \psi \sin 2 \theta_H \sin \phi_H~.
\eeq

In the helicity basis, the amplitudes $A_\parallel$ and $A_\perp$ are treated
on an equal footing, and the similarity of acceptances for these two amplitudes
then becomes manifest by the same dependence of $g_2$ and $g_3$ on $\psi$ 
and $\theta_H$.  The three terms $|A_{\parallel}|^2g_2 + |A_{\perp}|^2g_3 +
{\rm Im}(A^*_{\parallel}A_{\perp})g_4$
in the angular distribution (\ref{eqn:angdep}) are invariant under $\cos
\phi_H\leftrightarrow \sin \phi_H$, $A^*_\parallel \leftrightarrow A_\perp$.
As the angle $\phi_H$ describes only the relative orientations of the decay
planes defined by the decay products of $V_1 = J/\psi$ and $V_2 = \phi$, one
expects acceptance to depend weakly upon $\phi_H$, and to be similar for
the amplitudes $A_\parallel$ and $A_\perp$.  The acceptance for $A_0$ will
be {\it different} from that for $A_\parallel$ and $A_\perp$.  We advocate
attempting to extract the maximum information from the latter two amplitudes,
even though they account for less than half the decay
intensity. (See Table \ref{tab:comp} 
and corresponding values of $|A_0|^2$ and $|A_\parallel|^2$
in $B_s\to J/\psi\phi$ obtained in Refs.\,\cite{Abazov:2008fj,Aaltonen:2007gf}.)

Instead of Eq.~(\ref{angles+time}) the angular dependence is now given 
by the helicity functions $g_j(\vrp)$, while the time dependence involves 
the same transversity amplitudes $A_i$ and time-dependent functions,
${\cal T}_{\pm}, {\cal U}$ and  ${\cal V}$,  
$$
\frac{d^4 \Gamma [B_s(\obs) \to (\ell^+\ell^-)_{J/\psi} (K^+ K^-)_{\phi}]}
{d \cos \theta_H~d \phi_H~d \cos \psi~dt} \propto \frac{9}{32 \pi} \{
[|A_0|^2 g_1(\vrp) +|A_\parallel|^2 g_2(\vrp)]{\cal T}_+
 + |A_{\perp}|^2 g_3(\vrp)~{\cal T}_-
$$
\beq
+ |A_\parallel||A_{\perp}| g_4(\vrp)~{\cal U} + |A_0||A_\parallel|
\cos(\delta_\parallel) g_5(\vrp)~{\cal T}_+ + |A_0||A_{\perp}| g_6(\vrp)~
{\cal V} \}~~.
\eeq
To guard against systematic errors associated with incompletely understood
acceptances, we suggest weighting the angular distribution by
a function which emphasizes its dependence on $A_\parallel$ and $A_\perp$ and
de-emphasizes its dependence on $A_0$.  Such a function is
\beq
g(\cos \psi) = 3 - 5 \cos^2 \psi~,~~~~
\int_{-1}^1 d(\cos \psi) g(\cos \psi) \left\{ \begin{array}{c} \cos^2 \psi \cr
\sin^2 \psi \cr \sin 2\psi \end{array} \right\} = \left\{ \begin{array}{c} 0 \cr \frac{8}{3}
\cr 0 \end{array} \right\}~.
\eeq
Consequently, $\int_{-1}^1 d(\cos \psi) g(\cos \psi) g_j(\vrp)$ is
non-zero only for $j = 2,3,4$, corresponding to the three contributions 
involving only $A_\parallel$ and $A_\perp$. 

Integrating also over $\theta_H$ one obtains
$$
\int_{-1}^1 d(\cos \psi) g(\cos \psi) \frac{d^3 \Gamma [B_s(\obs) \to (\ell^+\ell^-)
_{J/\psi} (K^+ K^-)_{\phi}]} {d \phi_H~d \cos \psi~dt} \propto
$$
\beq 
\frac{1}{\pi} \left[|A_\parallel|^2\,(\frac{3}{2} - \cos\phi_H^2)\,{\cal T}_+ 
+ |A_{\perp}|^2\,(\frac{3}{2} - \sin\phi_H^2)\,{\cal T}_-
- |A_\parallel||A_{\perp}|\,\sin 2\phi_H\,{\cal U}\right]~.
\eeq
The three trigonometric functions of $\phi_H$ are linearly independent. 
This enables separate measurements of $|A_\parallel|, |A_\perp|$ and of
the time-dependent functions, ${\cal T}_+, {\cal T}_-$ and
${\cal U}$ for tagged $B_s$ ($\eta=+1$) and $\bar B_s$ ($\eta=-1$). This may be
used to determine $\beta_s$ and $\Delta\Gamma$.

\begin{figure}[h]
\begin{center}
\includegraphics[width=0.7\textwidth]{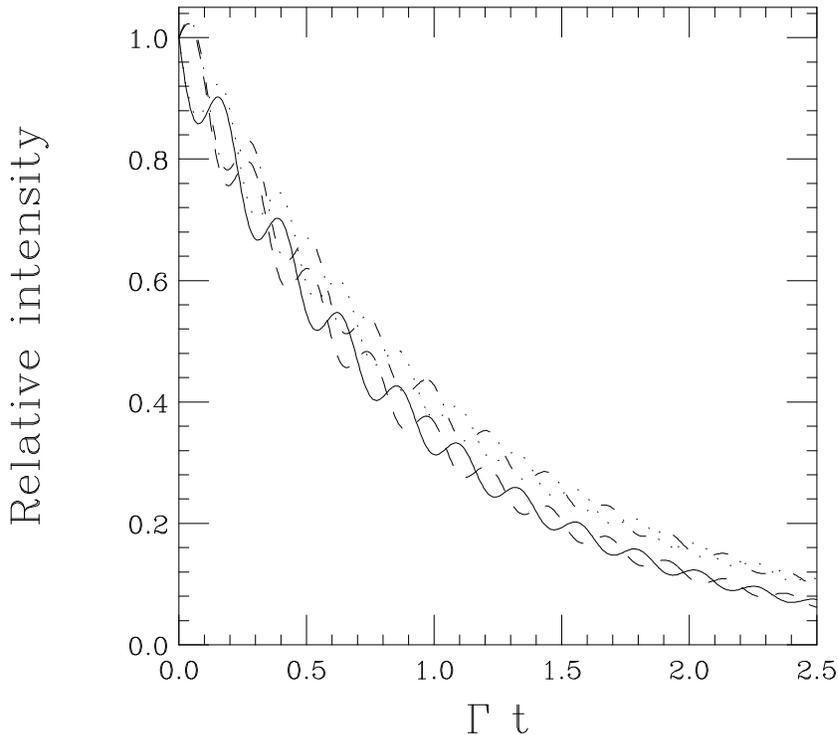}
\end{center}
\caption{Time-dependence of ${\cal T}_\pm$ for $B_s$ and $\obs$ tags based
on ``best-fit'' parameters for $B_s \to J/\psi \phi$ decays from Ref.\
\cite{Aaltonen:2007he} (see text).  Solid:  ${\cal T}_+$, $B_s$ tag.
Dashed:  ${\cal T}_+$, $\obs$ tag.  Dot-dashed:  ${\cal T}_-$, $B_s$ tag.
Dotted:  ${\cal T}_-$, $\obs$ tag.
\label{fig:T}}
\end{figure}
\bigskip

(2) Integration of angular distributions with all angles except the
transversity polar angle $\theta_T$ was shown in Ref.\ \cite{Dighe:1995pd}
to separate CP-even contributions, behaving as $1 + \cos^2 \theta_T$,
from CP-odd contributions, behaving as $\sin^2 \theta_T$.  Dependence upon
strong phases, contained only in interference terms which integrate to zero,
then disappears.  Important information is still retained in the functions
${\cal T}_\pm$, whose oscillatory behavior provides information on
$\beta_s$.  If $\sin 2 \beta_s$ is appreciable, as in the most recent fits
to CDF and D0 data based on tagged $B_s$ samples \cite{Aaltonen:2007he,%
Abazov:2008fj}, this time-dependence should be clearly visible.  As one
example, we have chosen ``best-fit'' values $\Delta \Gamma / \Gamma = 0.228$,
$2 \beta_s = 0.77$ from Ref.\ \cite{Aaltonen:2007he}, assumed a pessimistic
dilution factor $D = 0.11$ \cite{Aaltonen:2007he} to multiply the factor
$\eta$ in Eq.\ (\ref{eqn:T}), and display the functions ${\cal T}_\pm$ for
$B_s$ and $\obs$ tags.  The results are shown in Fig.\ \ref{fig:T}.
If the oscillations are visible in fits to data, there should be no question
about the observation of large $\sin 2 \beta_s$.  This method avoids possible
biases involving the extraction of $\beta_s$ from the interference terms
${\cal U}$ and ${\cal V}$ involving strong phases.

{\em Note added}: A month after the submission of this Letter the DO Collaboration 
reported new measurements of transversity amplitudes in $B_s\to J/\psi\phi$ and 
$B^0\to J/\psi K^*$ (assuming $\beta_s=0$ in the former) with values similar to 
the CDF measurements quoted in Eq.~(\ref{magnitudes})~\cite{Abazov:2008jz}:
\beq
|A_0| =  \left\{ \begin{array}{c} 0.745\pm 0.019,~B_s\to J/\psi\phi~~\cr 
0.766 \pm 0.011,~B^0\to J/\psi K^*\end{array}\right.,~~~
|A_\parallel| =  \left\{ \begin{array}{c} 0.494\pm 0.035,~B_s\to J/\psi\phi~~\cr 
0.480 \pm 0.029,~B^0\to J/\psi K^* \end{array} \right.~.
\eeq
This supports our assumption that the strong phase differences $\delta_\parallel$ 
and $\delta_\perp$ in these two processes are equal within $10^\circ$. 

M.G. would like to thank the Enrico Fermi Institute at 
the University of Chicago for its kind and generous hospitality. We thank 
J. Boudreau, K. Gibson, G. Giurgiu, C. Liu, and M. Paulini for 
useful discussions.  This work was supported in part by 
the United States Department of Energy through Grant No.\ DE FG02 90ER40560.

\end{document}